\DocumentMetadata{pdfstandard=A-2b}
\documentclass[conference]{IEEEtran}

\usepackage{cite}
\usepackage{graphicx}
\usepackage{url}
\usepackage{subcaption}
\usepackage{listings}
\usepackage{booktabs}
\usepackage{multicol}
\usepackage[multiple]{footmisc}
\usepackage{flushend}
\usepackage{balance}
\usepackage{hyperref}

\newcommand{\revealing}[1]{#1}

\newcommand{\q}[1]{``#1''}
\newcommand{\code}[1]{\texttt{#1}}

\newcommand{\ie}{i.e.,}

\lstdefinelanguage{domainmodel}{
	morekeywords={component,useCase,serviceCandidate,local,remote,asymmetric,serviceCandidate,dataStore,entityType},
	sensitive=true,
	morecomment=[l]{//},
}

\begin{document}

\title{Fast and Efficient What-If Analyses of\\ Invocation Overhead and Transactional Boundaries\\ to Support the Migration to Microservices}

\author{\IEEEauthorblockN{\revealing{Holger Knoche}}
\IEEEauthorblockA{\textit{\revealing{Software Engineering Group}} \\
\textit{\revealing{Kiel University}}\\
\revealing{Kiel, Germany} \\
\revealing{hkn@informatik.uni-kiel.de}}
\and
\IEEEauthorblockN{\revealing{Wilhelm Hasselbring}}
\IEEEauthorblockA{\textit{\revealing{Software Engineering Group}} \\
\textit{\revealing{Kiel University}}\\
\revealing{Kiel, Germany} \\
\revealing{wha@informatik.uni-kiel.de}}
}

\maketitle

\begin{abstract}
Improving agility and maintainability are common drivers for companies to adopt a microservice architecture for their existing software systems.
However, the existing software often relies heavily on the fact that it is executed within a single process space.
Therefore, decomposing existing software into out-of-process components like microservices can have a severe impact on non-functional properties, such as overall performance due to invocation overhead or data consistency.
	
To minimize this impact, it is important to consider non-functional properties already as part of the design process of the service boundaries.
A useful method for such considerations are what-if analyses, which allow to explore different scenarios and to develop the service boundaries in an iterative and incremental way.
Experience from an industrial case study suggests that for these analyses, ease of use and speed tend to be more important than precision.
	
In this paper, we present emerging results for an approach for what-if analyses based on trace rewriting that is (i) specifically designed for analyzing the impact on non-functional properties due to decomposition into out-of-process components and (ii) deliberately prefers ease of use and analysis speed over precision of the results.
\end{abstract}

\section{Introduction}

\noindent Agility is a key ability in modern software development.
Large, coordinated releases every few months cannot keep up with the needs of quickly delivering features to the end user, fixing bugs, or addressing security flaws.
An architectural style that is commonly considered an enabler for agility are microservices \cite{LewisFowler2014}.
This style is furthermore seen as viable for improving the maintainability of existing software as well as the time-to-market for new features \cite{TaibiLenarduzziPahl2017,KnocheHasselbring2019}.
Despite numerous benefits, however, microservice architectures come at a price; in parti\-cu\-lar, with respect to operational complexity and the required skills of the development and operations teams \cite{Fowler2015,TaibiLenarduzziPahl2017}.
In some cases, projects that have started using microservices have even moved to a monolithic architecture \cite{MendoncaEtAl2021}.

When decomposing an existing application into microservices, it is crucial to have viable service boundaries.
Service boundaries at the wrong locations can lead to severe problems~\cite{TaibiLenarduzziPahl2020}, especially regarding non-functional properties that may only surface in production.
In the worst case, the result is a \emph{distributed monolith}~\cite{Newman2020}, a system that has the disadvantages of both monoliths and microservices, but the advantages of neither.
Unlike service implementations, service boundaries are difficult to change in a microservice architecture, as they require multiple services to be changed in a coordinated way.

Based on observations from an industrial case study (see below), we focus on two non-functional properties that may indicate bad locations for service boundaries, namely invocation overhead and mismatched transaction boundaries.
The first may lead to low performance due to fine-grained excessive communication, while the latter may result in data inconsistencies or deadlocks.

Finding viable service boundaries is often an iterative and explorative process.
In this paper, we present emerging results for an approach to support this process by what-if analyses based on rewriting execution traces captured from the existing system.
To our knowledge, there is currently no other microservice decomposition approach that employs trace rewriting.
Since minor issues can often be remediated by additional hardware resources, we focus on major issues that justify or even require special treatment, such as adjusting service boundaries or implementing mitigation patterns.
We argue that for this purpose, an approximate prediction based on quantitative data from the existing application is sufficient, and that it is more important to have an approach that is easy to understand and use by practicioners.
Furthermore, the analysis should be fast enough to allow for interactive experimentation, at least for specific cases.

\section{Background / Industrial Case Study}
\noindent This work is based on an industrial case study, namely the decomposition of the core software system of an insurance company in Germany.
Development of this software began about forty years ago, and now consists of more than thirty applications.
Initially, the applications were implemented in COBOL on mainframe computers, while newer applications have been implemented in Java EE on Linux servers.

In order to improve agility, it was decided to move from scheduled, coordinated releases to individual releases of the applications on demand.
This, however, first required enabling the existing applications to be deployed independently.
A crucial step of this enabling is to move parts of the implementation to out-of-process components.
Especially the COBOL implementation is built on the assumption that multiple applications run in the same process, and therefore, (i) invocation overhead is negligible and (ii) everything runs in a single ACID transaction.
Decomposition into out-of-process components breaks these assumptions, and may negatively affect the non-functional properties of the system.
Although the Java applications are already distributed, they also make extensive use of (distributed) transactions.
As such transactions inhibit independent deployment and are commonly discouraged in microservice architectures \cite{Newman2021}, removing them is an important part of the decomposition process \cite{VernonJaskula2022}.

Due to the potential performance impact, there were concerns regarding the decomposition of our case study, especially with respect to large-volume batch operations that have to complete within a set time frame.
To investigate these concerns, it proved immensely helpful to quantify the invocations of the to-be services within the existing applications.
For this purpose, special monitoring function calls were inserted into the batch jobs to count and measure the duration of each function to be moved, and output a short summary at the end of each job.
Albeit simple, this data proved to be a great help in identifying critical invocations as well as preparing countermeasures and optimizations.
Furthermore, it provided an important link to the business domain by showing the effect of the decomposition on business-related elements, in this case, important batch jobs.
Finally, it provided valuable insights for the developers, who were aware of the operations immediately invoked by their programs, but not their transitive dependencies.
The data was, however, insufficient to analyze potential data consistency issues.

\section{Proposed Approach}
\label{sec:approach}

\noindent Based on the experience from our case study, we propose the following approach to perform what-if analyses to identify bad locations for service boundaries.
First, we describe prerequisites for its application, and give a short overview.
Then, we present more details on key elements of the approach, namely deployment models, trace rewriting and trace analysis.

\subsection{Prerequisites and Overview}
\noindent Since our approach aims at analyzing and improving service boundaries, we assume that at least a draft of these boundaries already exists.
We require the service boundaries to be specified as a set of \emph{service candidates}, \ie{} existing functionality that is to be provided as service operations, which map to invokable entities within the existing software system.
Such entities might be functions, methods, or programs, depending on the programming language.
We furthermore assume that the relevant use cases and data entity types are known and can be associated with the existing implementation.

\begin{figure*}[bt]
  \begin{center}
    \includegraphics[width=0.95\linewidth]{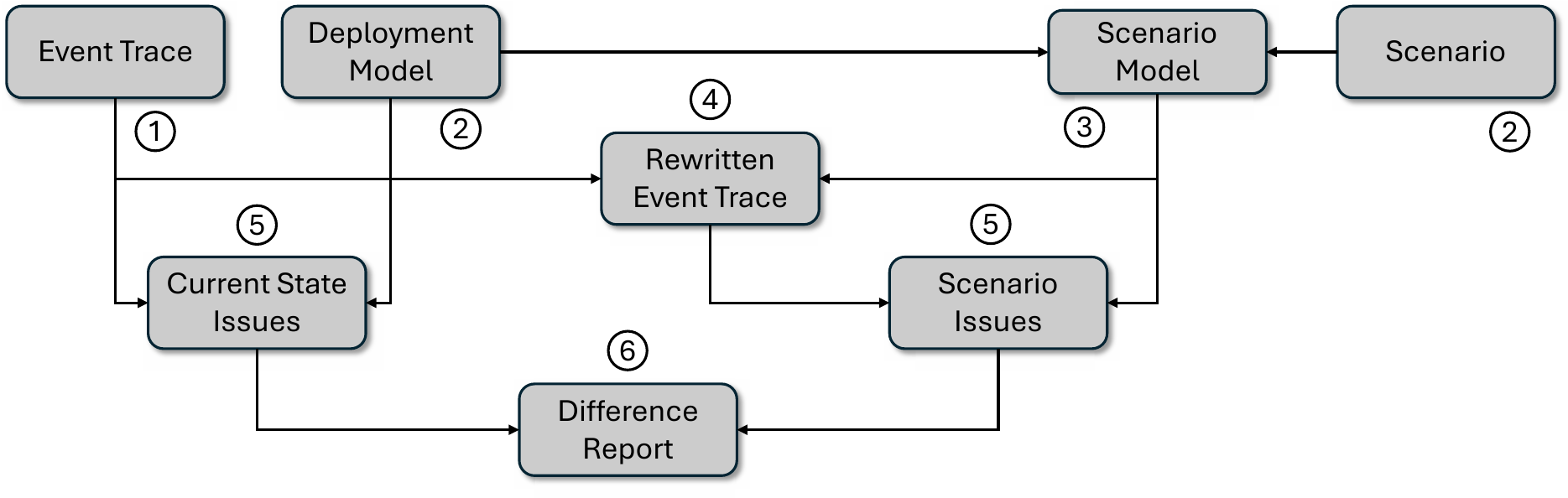}
    \end{center}
    \caption{Overview of our Approach. The circled numbers indicate the steps in which the respective element is used.}
    \label{fig:approach-overview}
\end{figure*}

Our approach can be roughly structured into the following steps, which are also depicted in Figure~\ref{fig:approach-overview}.
First, we observe the existing software in production and collect relevant events at runtime, especially those regarding invocations of the service candidates.
Events that occur in a coherent execution of a use case form an \emph{event trace} (Step~1 in Figure~\ref{fig:approach-overview}).
The assignment of traces to use cases is important for establishing a connection to the business domain.

In addition to the traces, we require one or more \emph{scenarios} to be analyzed.
Each scenario assigns the service candidates, use cases, and entity types to \emph{(service) components}, and describes how these components are intended to be deployed relative to each other.
This means that we do not model the deployment explicitly in terms of nodes and network connections, but only specify \emph{connections} between components that may be either \emph{local} (\ie{} the components reside in the same process) or \emph{remote}.
Since scenarios can often be specified more easily as deltas to the current deployment, we furthermore require a model of the current state of the application (Step 2).
This provides us with the benefit that the current state and the scenario can be analyzed in exactly the same way.

The next step of our approach is the actual what-if analysis, which -- given event traces, a current deployment model, and a scenario -- is performed as follows.
First, the scenario and the current deployment model are combined to derive the scenario deployment model (Step~3).
Then, the existing traces are rewritten with respect to overhead and transactional boundaries to match the scenario deployment model (Step 4).
All other aspects are left unchanged (\emph{ceteris paribus}).
As part of the rewriting process, a mapping between the rewritten and original events is established.

After rewriting, both the original and the rewritten traces are analyzed with respect to invocation overhead, duration, potential consistency issues, and committed and reverted writes (Step~5 in Figure~\ref{fig:approach-overview}).
Since both the current state and the scenario state are represented as event traces with a matching deployment model, the same analyses can be used.

Finally, the results from both analyses are compared to highlight changes that are induced by the scenario (Step 6).
The event mapping plays an important role in this task, as it is required to identify issues that appear in both states.
For the overhead analysis, we are particularly interested in (statistically) significant changes in duration, whereas the consistency analysis focuses on changes in potential consistency issues as well as writes to entities that change in terms of being committed or reverted.
The results are then presented to the designer in appropriate ways; in our proof-of-concept implementation, we use graphical visualizations of the traces as well as tables.

We envision these what-if analyses to be used in an iterative way.
Starting with an initial set of service boundaries, the necessary traces are gathered from the existing system and analyzed for different scenarios.
If sufficiently severe issues are identified, appropriate mitigation stategies can be selected, guided by actual quantitative data from production.
The analysis may, however, also reveal issues that prevent the desired strategy from being applied immediately, such as consistency requirements that can no longer be satisfied.
In this case, the implementation can be changed to remove these impediments, resulting in a new iteration.
The same applies if additional service candidates are identified.
This process continues until the developers are confident that all relevant issues have been addressed, and the decomposition can actually be implemented.

\subsection{Deployment Models}
\noindent Our approach uses deployment models to specify the current and intended deployment in an abstract way.
In the following paragraphs, we briefly sketch the structure of such a model.
The central element of such a model is a \emph{component}.
Components conceptually represent a group of \emph{use cases}, \emph{service candidates}, and \emph{entity types} that are deployed together such that (i) there is negligible invocation overhead and (ii) transactions are shared across invocations.
Usually, this implies that a component runs within a single process space with a single database.

Components can have \emph{component connections} to other components, which can be either \emph{local} or \emph{remote}.
Local connections express that the respective components are deployed together, and that invocations between them have the same properties as internal invocations.
Remote connections, on the other hand, incur invocation overhead and are restricted to using distributed transactions or being unable to propagate transactions.
A component (or group of components connected only by local connections) is considered a potential microservice if none of its incoming and outgoing remote connections propagates transactions.

Listing 1 shows an example model of a fictional car insurance, specified in a simple textual language.
It consists of two components connected by a local connection.
This example also shows that service candidates can express their transaction behavior using the well-known transaction attributes from Jakarta EE (in this case, \code{REQUIRED} to create a new transaction if necessary).
Furthermore, an entity type is defined and assigned to a data store, which specifies the behavior in case of a read-write conflict.
In the example, the default is used, which is to return possibly stale data.

\newpage
    \small
    \begin{lstlisting}[language=domainmodel, caption=Example Domain Model Specification]
  component "Car Insurance" {
    useCase "Create Car Contract"
    serviceCandidate createCarContract [
      transactionBehavior = REQUIRED
    ]
    entityType CarContract
  }
  component "Contracts" {
    serviceCandidate createContract
  }
  local "Car Insurance" -> "Contracts"
  dataStore "Shared Database" {
    entityType CarContract
  }
    \end{lstlisting}
    \normalsize

As previously noted, scenarios are specified as deltas to the current state.
Listing 2 shows an example of a scenario specification, which changes the transaction behavior of the \code{createContract} candidate to \code{REQUIRED} and the formerly local component connection to a remote one with 10 units (for instance, milliseconds) of latency and default transaction propagation, which is no propagation.

    \small
    \begin{lstlisting}[language=domainmodel, caption=Example Scenario Specification]
  component "Contracts" {
    serviceCandidate createContract [
      transactionBehavior = REQUIRED
    ]
  }
  remote "Car Insurance" -> "Contracts" [
    overhead = 10
  ]    
    \end{lstlisting}
    \normalsize

\subsection{Trace Rewriting}
\noindent In order to analyze the effects of a scenario on the runtime behavior, we rewrite the observed traces as if they had occurred in this scenario.
For our approach, we perform two types of rewriting: \emph{overhead rewriting} and \emph{transaction boundary rewriting}.
In both cases, we follow the trace through the deployment model and apply modifications when encountering changes introduced by a scenario.

Rewriting with respect to invocation overhead is mostly concerned with adjusting timestamps.
During monitoring, overhead is captured as the time difference between the invocation of a service candidate and the entry into its implementation; the same applies to the return from a candidate.
Once an invocation-entry pair is encountered in the trace, the component of the invoked service candidate and the connection between the current and invoked component are determined from the scenario model.
If this connection was modified or created by the scenario, the timestamp of the entry event is adjusted so that the difference matches the overhead of the connection, where local connections have, by definition, an overhead of zero.
The timestamps of subsequent events are modified accordingly.

Rewriting transaction boundaries is considerably more complex.
Although this rewriting only affects the flag of whether or not a transaction was started on entering a service candidate, determining the correct value requires to keep track of the transaction state during the entire trace.
We therefore fully emulate the transaction management by the container as well as remote context propagation and two-phase commits where applicable.

\subsection{Trace Analysis and Visualization}
\noindent To identify possible issues due to the decomposition, we perform several analyses on both the original and the rewritten traces.
As for potential issues due to invocation overhead, we first rewrite the input traces as described above.
Then, we calculate the duration of each input trace and its rewritten counterpart, and perform a significance test on the traces for each individual use case.
For this purpose, we use a two-sample Welch test, a modified $t$-test applicable to samples with different variances.
A paired $t$-test proved to be impractical as it considered even tiny changes as significant.
This allows us to identify use case cases that are significantly impacted by the analyzed scenario, and therefore require further attention.

As for transaction boundaries, we are particularly interested in potential consistency issues that might inhibit the application of mitigation strategies, or require special treatment.
For instance, many patterns to reduce the impact of invocation overhead rely on parallel or asynchronous processing, and may therefore break order-related consistency expectations such as \q{read-your-writes} consistency~\cite{TerryEtAl1994}.
An indicator for the violation of such expectations are access conflicts on entities that only appear in the rewritten trace.
Another example is the saga pattern~\cite{GarciaMolina1987}, which decomposes a large transaction into a sequence of shorter transactions.
This decomposition may require to explicitly compensate changes that would previously have been reverted automatically.
The need for such compensation is indicated by writes to entities that are reverted in the original trace, but are committed in the rewritten trace.

In order to foster interactive experimentation, we also propose a way to incorporate additional context information like transaction boundaries into trace visualizations.
Many current observation tools use the notion of a trace as described by OpenTelemetry.\footnote{\url{https://opentelemetry.io/docs/concepts/signals/traces/}}
In a nutshell, such a trace consists of \emph{spans}, which represent a specific operation or another action that has a duration.
Spans can be arranged in a hierarchy to mirror the call hierarchy, and can contain \emph{span events} which occur at a specific instant in time.
For trace visualization, the spans are usually depicted as bars on a common timeline.

We extend this visualization by adding \emph{span overlays}.
A span overlay is associated with a span, but may extend over either end of the span itself.
This allows to depict both overhead and transactions as specific overlays.

\begin{figure}
  \begin{center}
      \includegraphics[width=\linewidth]{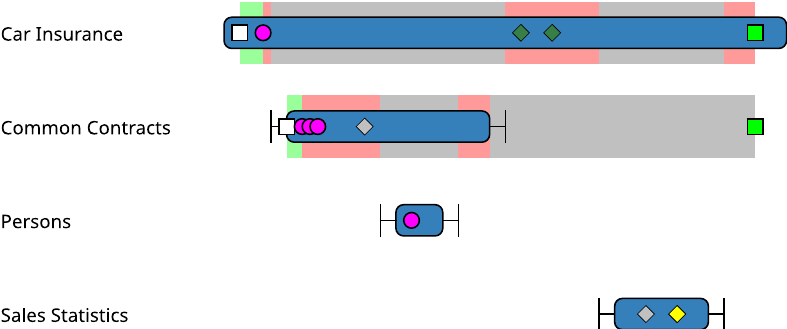}
      \end{center}
      \caption{Example Trace Visualization. The blue bars represent the spans, and the small shapes within the bars represent different types of events:
Rectangles represent transaction events (in the example, start (white) and commit (green) of a transaction), circles represent accesses to entities, and diamonds represent consistency issues of different types. For instance, the yellow diamond represents a read of an entity with pending changes in another transaction, which may indicate an expectation of read-your-writes consistency. Invocation overhead is depicted as the \q{whiskers} before and after the respective span.
The colored bars in the background represent the transaction state at the corresponding point in time:
A green bar denotes that a transaction is active and has no pending writes, whereas a red bar indicates an active transaction with pending writes.
The time during which a transaction is suspended is represented by a gray bar.
}
      \label{fig:example-trace}
\end{figure}

Consider the example shown in Figure~\ref{fig:example-trace}, which shows the trace of hypothetical contract creation.
Note that the transaction overlay of the common contracts invocation extends over the end of the corresponding span, as it represents a subordinate transaction that is committed atomically with its top-level transaction.

\section{Preliminary Results}
\noindent As of now, we have created a proof-of-concept implementation of the trace rewriting, analysis, and visualization.
Experiments with artificial event traces not only show the overall viability of the approach, but also indicate that the analysis is fast enough to process substantial data sets in a reasonable amount of time.
To give an impression, the analysis of a single trace with 2.5 million remote invocations took about 2.4 seconds on a current desktop PC with an Intel Core i7-14700F processor, and the analysis of a set of 100,000 random traces with at most 50 remote invocations per trace took about 2.0 seconds on one core, and about 0.4 seconds utilizing all 20 cores.

We furthermore conducted a survey with seven professional software developers and six students to investigate the understandability of our approach, its perceived relevance in practice, and the usefulness of specific aspects.
Overall, the results were very positive, indicating that our approach addresses a relevant problem, and fulfills our goal of being understandable by practicioners.
An aspect that was particularly well received was the focus on changes induced by a scenario, which was greatly facilitated by using trace rewriting as the underlying method.

\section{Discussion}
\noindent As noted in the introduction, we are not aware of any design-time analysis approach for microservice migration that uses trace rewriting as its underlying method.
Therefore, we can currently only rely on our own experiments to assess the viability of this method.
Although the preliminary results are promising, a major limitation comes from the fact that until now, experiments have only been performed on artificial data, not data from a real application.
Acquiring such data, however, is not trivial:
While monitoring tools such as Kieker \cite{HasselbringVanHoorn2020} are readily available and able to capture many of the required events, they currently offer little support for capturing transaction-related and entity-related events.

Data from real applications will also provide experimental evidence to support our initial hypothesis that an approximate prediction of the performance impact is sufficient to identify bad locations for service boundaries.
Our survey participants, however, already considered the results as sufficiently precise.

A fundamental limitation of approaches based on dynamic analysis is that they can only analyze what is actually executed.
Therefore, use cases that occur seldomly may not be analyzed since they were not executed during the observation period, or too few observations may be available for sound conclusions.
In our case study, some tasks are only executed once a year, which does not only imply a large observation period, but also that a considerable amount of data must be kept for a long time.

As for alternatives to our approach, there is a plethora of approaches for decomposing existing software into microservices; 
literature surveys include \cite{FritzschEtAl2019}, \cite{CapuanoMuccini2022}, and \cite{MartinezEtAl2025}.
Most of the existing approaches, however, do not address the issue of transactions:
The last of the mentioned studies finds that although several of the reviewed publications were concerned with the database migration, only one of them addressed the issue of transactions.
We therefore consider our approach complementary to existing approaches by addressing this particular issue.

More sophisticated performance models, such as the Palladio Component Model~\cite{ReussnerEtAl2016}, are considerably more powerful and versatile.
These properties, however, come at the cost of a significantly higher model complexity, which is detrimental to our goal of being easy to use and understand for practicioners.

\section{Future Plans}
\noindent Our primary goal for the future is to further evaluate and mature our approach.
As the next step, we therefore plan to apply it to real applications, such as common benchmark applications as well as our industrial case study.
This will require to implement an appropriate monitoring solution, most likely by extending Kieker to capture the necessary events.
We furthermore intend to extend our approach to detect additional opportunities for potential optimizations, such as service candidates that are always invoked together and could be combined into a single one.
The trace visualization also provides opportunities for future work, especially for working with large traces.

During our survey, we received feedback that our prototype helped some of the participants to better understand the side-effects of service and transaction boundaries.
We therefore consider its use in developer education to be another potential opportunity for future work.

\providecommand{\doi}[1]{DOI: \href{https://doi.org/#1}{#1}}


\begin{thebibliography}{10}
\providecommand{\url}[1]{#1}
\csname url@samestyle\endcsname
\providecommand{\newblock}{\relax}
\providecommand{\bibinfo}[2]{#2}
\providecommand{\BIBentrySTDinterwordspacing}{\spaceskip=0pt\relax}
\providecommand{\BIBentryALTinterwordstretchfactor}{4}
\providecommand{\BIBentryALTinterwordspacing}{\spaceskip=\fontdimen2\font plus
\BIBentryALTinterwordstretchfactor\fontdimen3\font minus
  \fontdimen4\font\relax}
\providecommand{\BIBforeignlanguage}[2]{{%
\expandafter\ifx\csname l@#1\endcsname\relax
\typeout{** WARNING: IEEEtran.bst: No hyphenation pattern has been}%
\typeout{** loaded for the language `#1'. Using the pattern for}%
\typeout{** the default language instead.}%
\else
\language=\csname l@#1\endcsname
\fi
#2}}
\providecommand{\BIBdecl}{\relax}
\BIBdecl

\bibitem{LewisFowler2014}
\BIBentryALTinterwordspacing
J.~{L}ewis and M.~{F}owler. (2014) {M}icroservices. [Online]. Available:
  \url{http://martinfowler.com/articles/microservices.html}
\BIBentrySTDinterwordspacing

\bibitem{TaibiLenarduzziPahl2017}
D.~Taibi, V.~Lenarduzzi, and C.~Pahl, ``{P}rocesses, {M}otivations, and
  {I}ssues for {M}igrating to {M}icroservices {A}rchitectures: {A}n {E}mpirical
  {I}nvestigation,'' \emph{IEEE Cloud Computing}, vol.~4, no.~5, pp. 22--32,
  2017.

\bibitem{KnocheHasselbring2019}
H.~Knoche and W.~Hasselbring, ``{D}rivers and {B}arriers for {M}icroservice
  {A}doption -- {A} {S}urvey among {P}rofessionals in {G}ermany,''
  \emph{International Journal of Conceptual Modeling}, vol.~14, no.~1, pp.
  1--35, 2019.

\bibitem{Fowler2015}
\BIBentryALTinterwordspacing
M.~Fowler. (2015) Microservice {P}remium. [Online]. Available:
  \url{https://martinfowler.com/bliki/MicroservicePremium.html}
\BIBentrySTDinterwordspacing

\bibitem{MendoncaEtAl2021}
N.~C. Mendonça, C.~Box, C.~Manolache, and L.~Ryan, ``The {M}onolith {S}trikes
  {B}ack: {W}hy {I}stio {M}igrated {F}rom {M}icroservices to a {M}onolithic
  {A}rchitecture,'' \emph{IEEE Software}, vol.~38, no.~5, pp. 17--22, 2021.
  \doi{10.1109/MS.2021.3080335}

\bibitem{TaibiLenarduzziPahl2020}
D.~Taibi, V.~Lenarduzzi, and C.~Pahl, ``{M}icroservices {A}nti-{P}atterns: {A}
  {T}axonomy,'' in \emph{Microservices: Science and Engineering}.\hskip 1em
  plus 0.5em minus 0.4em\relax Springer, 2020.
  \doi{10.1007/978-3-030-31646-4\_5} pp. 111--128.

\bibitem{Newman2020}
S.~Newman, \emph{Monolith to Microservices: Evolutionary Patterns to Transform
  Your Monolith}.\hskip 1em plus 0.5em minus 0.4em\relax O'Reilly, 2020.

\bibitem{Newman2021}
------, \emph{Building Microservices: Designing Fine-grained Systems}.\hskip
  1em plus 0.5em minus 0.4em\relax {O}'{R}eilly, 2021.

\bibitem{VernonJaskula2022}
V.~Vernon and T.~Jaskula, \emph{Strategic Monoliths and Microservices}.\hskip
  1em plus 0.5em minus 0.4em\relax Pearson, 2022.

\bibitem{TerryEtAl1994}
D.~Terry, A.~Demers, K.~Petersen, M.~Spreitzer, M.~Theimer, and B.~Welch,
  ``{S}ession {G}uarantees for {W}eakly {C}onsistent {R}eplicated {D}ata,'' in
  \emph{Proc. 3rd Int. Conf. on Parallel and Distributed Information Systems},
  1994. \doi{10.1109/PDIS.1994.331722} pp. 140--149.

\bibitem{GarciaMolina1987}
H.~Garcia-Molina and K.~Salem, ``Sagas,'' \emph{ACM SIGMOD Record}, vol.~16,
  no.~3, pp. 249--259, 1987. \doi{10.1145/38714.38742}

\bibitem{HasselbringVanHoorn2020}
W.~Hasselbring and A.~{van Hoorn}, ``{K}ieker: {A} {M}onitoring {F}ramework for
  {S}oftware {E}ngineering {R}esearch,'' \emph{Software Impacts}, vol.~5, pp.
  1--5, 2020. \doi{10.1016/j.simpa.2020.100019}

\bibitem{FritzschEtAl2019}
J.~Fritzsch, J.~Bogner, A.~Zimmermann, and S.~Wagner, ``{F}rom {M}onolith to
  {M}icroservices: {A} {C}lassification of {R}efactoring {A}pproaches,'' in
  \emph{Software Engineering Aspects of Continuous Development and New
  Paradigms of Software Production and Deployment}.\hskip 1em plus 0.5em minus
  0.4em\relax Springer, 2019. \doi{10.1007/978-3-030-06019-0\_10} pp. 128--141.

\bibitem{CapuanoMuccini2022}
R.~Capuano and H.~Muccini, ``{A} {S}ystematic {L}iterature {R}eview on
  {M}igration to {M}icroservices: {A} {Q}uality {A}ttributes {P}erspective,''
  in \emph{19th Int. Conf. on Software Architecture Companion (ICSA-C)}, 2022.
  \doi{10.1109/ICSA-C54293.2022.00030} pp. 120--123.

\bibitem{MartinezEtAl2025}
A.~{Martínez Saucedo}, G.~Rodríguez, F.~{Gomes Rocha}, and R.~P. dos Santos,
  ``{M}igration of {M}onolithic {S}ystems to {M}icroservices: {A} {S}ystematic
  {M}apping {S}tudy,'' \emph{Information and Software Technology}, vol. 177,
  2025. \doi{10.1016/j.infsof.2024.107590}

\bibitem{ReussnerEtAl2016}
R.~H. Reussner, S.~Becker, J.~Happe, R.~Heinrich, A.~Koziolek, H.~Koziolek,
  M.~Kramer, and K.~Krogmann, \emph{Modeling and Simulating Software
  Architectures: The Palladio Approach}.\hskip 1em plus 0.5em minus 0.4em\relax
  The MIT Press, 2016.

\end{thebibliography}
\end{document}